\documentclass[12pt,a4paper]{article}
\usepackage{bbm}
\usepackage{amsmath}
\usepackage{amssymb}
\usepackage{graphicx}
\usepackage{psfrag}
\begin{document}

\begin{flushright}
BI-TP 2000/38 \\
\end{flushright}

\begin{center}  
\large{\bf Geometry of Reduced Supersymmetric 4D Yang-Mills Integrals}
\end{center}

\begin{center}  
Z. Burda$^{1,2}$, B. Petersson$^{1}$ and J. Tabaczek$^{1}$
\end{center}

\centerline{$^{1}$Fakult\"at f\"ur Physik, Universit\"at Bielefeld} 
\centerline{P.O.Box 100131, D-33501 Bielefeld, Germany}
\vspace{0.3cm}
\centerline{$^{2}$Institute of Physics, Jagellonian University} 
\centerline{ul. Reymonta 4, 30-059 Krakow, Poland}
\vspace{0.3cm}

\begin{abstract} \normalsize \noindent
We study numerically the geometric properties of reduced supersymmetric
non-compact $SU(N)$ Yang-Mills integrals in $D = 4$ dimensions, for
$N = 2, 3, \dots, 8$. We show that in the range of large eigenvalues of
the matrices $A^\mu$, the original $D$-dimensional rotational symmetry 
is spontaneously broken and the dominating field configurations become
one-dimensional, as anticipated by studies of the underlying surface
theory. We also discuss possible implications of our results for
the IKKT model.
\end{abstract}

\section*{Introduction}

Reduced supersymmetric Yang-Mills integrals appear in many areas of
fundamental physics ranging from $QCD$ to string theory. 
An important issue is whether such reduced integrals are able to
capture universal properties of the corresponding supersymmetric 
Yang-Mills theory in the large $N$ limit, 
a question similar to the one posed by Eguchi
and Kawai in the context of standard $QCD$ \cite{ek,gk}.
Reduced supersymmetric Yang Mills integrals also enter calculations of the
Witten index \cite{s} of a wide class of supersymmetric quantum
mechanical systems in D-brane physics \cite{y,ss}. Finally, in recent
years a non-perturbative definition of string theory was proposed
in terms of a ten-dimensional reduced supersymmetric Yang-Mills
integral -- the IKKT model -- which is believed to reproduce, in an
appropriate large $N$ limit, amplitudes including all topological
contributions of IIB critical strings \cite{ikkt}.

As a supersymmetric model, the reduced Yang-Mills integral can be
defined in $D = 3, 4, 6$, or 10 dimensions. As it turns out, many
important properties of the model, such as the singular behaviour
of the field correlators, change with $D$ in a systematic way, so that
results gained for one value of $D$ can be extrapolated to
the other cases. We will in this paper focus on the four-dimensional
model. In this case, the effective action, obtained by
integrating out the fermionic degrees of freedom, 
is positive semi-definite which
makes it well-suited to Monte Carlo studies.

Analytically, the model -- in any dimension -- has so far been solved
only for $N = 2$ \cite{s,ss,su2}. 
For larger $N$, there exist some results from approximate analytic methods
\cite{1loop} and numerical simulations \cite{mc1,tail,mc2,mc2',phase}.
This made it possible to address questions related
to the properties of the eigenvalue spectrum \cite{mc1,tail} as well as
the scaling of physical quantities like the gyration radius, correlation
functions, and Wilson loops \cite{mc2,mc2',phase}.

In this paper, we will study the geometrical properties of the model
in more detail. Our motivation comes from earlier studies of the
geometrical structure exhibited by the corresponding surface theory
\cite{largeN,largeN'}. This model was examined numerically in a
discretized form, using the dynamical triangulation 
approach \cite{oy,bbpt}. The distribution 
of the gyration radius was found to have a power-like tail.

For the matrix model one expects the same kind of large $R$ behaviour.
More precisely, the analytic solution for $N = 2$ exhibits a tail
$\rho (R) \sim R^{-2D+5}$, and it was conjectured from results of the
one-loop approximation that this formula should also hold for any $N>2$.
Intriguingly, in the surface theory for $D=4$ one observes
$\rho(R) \sim R^{-3}$, which agrees with the result from the matrix
model. A mechanism was proposed to explain the appearance of this tail,
relying on the emergence of essentially one-dimensional configurations
called `needles' or `tubes'. Numerical simulations for $D=4$ were able
to show that such configurations do indeed dominate the large $R$ part
of the $\rho (R)$ distribution. If one assumes the dominance of the
`tubes' for any $D$, one finds exactly 
the same formula as in the matrix model,
$\rho (R) \sim R^{-2D+5}$.

An obvious question that arises at this point is whether the
tail that appears in the matrix model could be explained by a similar
mechanism, {\em i.~e.} whether the geometrical structure of large
eigenvalue configurations also becomes one-dimensional. Trying to answer
this question is the central aim of this paper. We will again use
numerical simulations to examine this point.

We begin the paper by shortly reviewing the matrix
model of reduced Yang-Mills integrals.
We then describe the observables that allow us to pick up the geometrical
structure of a given configuration, and present the numerical results.
We finish with a short discussion.

\section*{The model}

The zero-dimensional supersymmetric Yang-Mills integral is defined
in the Euclidean sector by the partition function \cite{ikkt}
\begin{equation}
Z = \int dA \ d\bar{\Psi} \ d\Psi \ e^{-S [A, \bar{\Psi}, \Psi]} \, ,
\label{ikkt}
\end{equation}
where the action is given by
\begin{equation}
S [A, \bar{\Psi}, \Psi] = S_B + S_F
= - \frac{1}{4} \ {\rm Tr} \ [ A^\mu, A^\nu ]^2 \
+ \ \frac{1}{2} \ {\rm Tr} \ \bar{\Psi} \Gamma_\mu [ A^\mu, \Psi ] \, .
\label{action}
\end{equation}
The model is believed to provide a non-perturbative definition of
string theory. The $N \times N$ traceless Hermitean matrices $A^\mu$,
$\mu = 1, \dots, D$, are a sort of quantum operators for the bosonic
coordinates $X^\mu(\tau,\sigma) \rightarrow A^\mu_{ij}$ 
of the string world-sheet in the $D$-dimensional target
space. The fields $\bar{\Psi}$,$\Psi$ are fermionic
matrices that transform as Majorana-Weyl spinors in $D = 10$ dimensions,
and as Weyl spinors in $D = 4$. The IKKT model corresponds to the
case $D = 10$.

As in any quantum theory, one is interested in measuring correlation
functions, the simplest of which are one--matrix correlators
of the type $\langle \frac{1}{N} ({\rm Tr} A_\mu^2)^k \rangle$. Because the
model is rotationally invariant, such correlators do not
depend on the choice of $A^\mu$. More specifically, for any matrix of the
form $A = n_\mu A^\mu$, where $n_\mu$ is a unit vector, the
correlators give identical results.

One-matrix correlators can be expressed as moments of the distribution
of eigenvalues of the $A^\mu$,
\begin{equation}
\left\langle \frac{1}{N} \, ({\rm Tr} A^2)^k \right\rangle
= \int d\lambda \ \rho (\lambda) \ \lambda^{2k} \, ,
\label{1corr}
\end{equation}
where
\begin{equation}
\rho (\lambda) = \left\langle \frac{1}{N} \sum_{n=1}^N
\delta (\lambda - \lambda_n) \right\rangle \, .
\end{equation}
For $N = 2$, the eigenvalue distributions can be found analytically
for $D = 4, 6$ and $10$ dimensions \cite{su2}, 
and were shown to have the large $\lambda$ behaviour
\begin{equation}
\rho(\lambda) \sim \lambda^{-2D+5} \, .
\label{lpower}
\end{equation}
For higher $N$ the integrals have not been solved analytically. However,
it was conjectured \cite{su2}, based on results of the one-loop
approximation \cite{1loop}, that the large $\lambda$ part of the spectrum
should be controlled by the same powers (\ref{lpower}) independently of $N$.
 
The power law (\ref{lpower}) describes exactly the same behaviour as was
found for the surface theory, in which case it was shown
that the dominating configurations are one-dimensional \cite{bbpt} 
with scaleless fluctuations of the extension in the
elongated direction.

One way of extracting the dimensionality of the surface model is to
measure the correlation matrix
\begin{equation}
C_{\mu\nu}  = \frac{ \int \sqrt{g} \ X_\mu X_\nu}{\int \sqrt{g}} \, .
\end{equation}
The trace of this matrix gives the square of the gyration radius,
\begin{equation}
R^2  = \sum_\mu C_{\mu\mu} = 
\frac{ \int \sqrt{g} \ \sum_\mu X_\mu^2}{\int \sqrt{g}} \, .
\end{equation}
We can simplify things by choosing a basis in which $C_{\mu\nu}$
becomes diagonal,
\begin{equation}
C_{ij} = \frac{1}{N} \, \delta_{ij} \, r_i^2 \, .
\end{equation}
In this case, $R^2 = \sum_i r_i^2$. Also, the individual eigenvalues
$r_i^2$ now describe the square extent of the system in the directions
given by this particular basis.

For large $R$, the distribution of $R$ was found to behave as
$\rho (R) \sim R^{-3}$, just as in (\ref{lpower}). Also, one
of the eigenvalues $r_i^2$ was shown to become much larger than
the others, reflecting the one-dimensional nature of the large $R$
configurations in the surface theory \cite{bbpt}.

In analogy to this, we can now try to determine the dimensionality
of the matrix model. We can define the quantum correlation matrix as
\begin{equation}
C_{\mu\nu} = \frac{1}{N} {\rm Tr} A_\mu A_\nu
\end{equation}
and the square of the gyration radius $R^2$ as
\begin{equation}
R^2 = \sum_\mu C_{\mu\mu} = \frac{1}{N} {\rm Tr} \sum_\mu A^2_\mu \, .
\end{equation} 
Given that we can freely rotate the $A^\mu$ as discussed above, we can
again arrange things such that we find a diagonal matrix $C_{ij}$, and
the gyration radius once more becomes $R^2 = \sum_i r^2_i$. The
eigenvalues $r_i^2$ are all real numbers and can again be interpreted as
the square extent of the system in the directions $i$, except that this
time they also include quantum fluctuations.

The effective dimensionality can now be determined from the distribution
of the $r^2_i$. If $d$ out of $D$ eigenvalues can be shown to be much
larger than the remaining ones, the system can be said to be effectively
$d$-dimensional.
 
Let us order the eigenvalues by size, $r^2_i \ge r^2_{i+1}$.
We will study the behaviour of $r^2_i$ as a function of $R^2$~: 
\begin{equation}
\langle r^2_i \rangle_{R^2} = 
\frac{ \int \ e^{-S [A, \bar{\Psi}, \Psi]} \ \delta (R^2 (A) - R^2 ) \
r^2_i(A)}{ \int e^{-S [A, \bar{\Psi}, \Psi]} \ \delta (R^2 (A) - R^2)}
\label{r2R2}
\end{equation}
>From what we found in the surface model, we expect for large $R$ to
see $\langle r^2_1 \rangle_{R^2} \sim R^2$ whereas the
other $\langle r^2_i \rangle_{R^2}$ should, in comparison,
become negligible. It is convenient to introduce a new observable
\begin{equation}
\eta = \frac{r^2_2+r^2_3+r^2_4}{R^2} = 1 - \frac{r^2_1}{R^2}
\end{equation}
as a quantitative measure of this asymmetry. It tells us which fraction
of the square extent of the system comes from the transverse directions.
If the system does indeed become one-dimensional for large $R$, we should
find $\langle \eta \rangle_{R^2} \rightarrow 0$ for $R^2 \rightarrow \infty$.

\section*{Computational method}

Using the chiral representation of the gamma matrices and replacing
the bispinors $\Psi$ by two-component spinors $\psi$
\begin{equation}
\bar{\Psi} = (0 , \bar{\psi}) \ , \quad
\Psi = \left(\begin{array}{c} \psi \\ 0 \end{array}\right)
\end{equation}
we can re-write the fermionic part of the action as
\begin{equation}
S_F = \frac{1}{2} \, \bar{\psi}^a_{ij} \, {\cal M}_{aij,bkn} \, \psi^b_{kn}
\end{equation}
where the matrix ${\cal M}$ is given by
\begin{equation}
{\cal M}_{aij,bkn} = A^\mu_{jk} \gamma_\mu^{ab}\delta_{ni} -
              A^\mu_{ni} \gamma_\mu^{ab}\delta_{kj}
\end{equation}
and the gamma matrices are $\gamma_\mu = (-i \sigma_0, \vec{\sigma})$.
Here, space-time indices are denoted by Greek letters, matrix indices
by $i, j, k, \dots$, and spinor indices  by $a, b, c, \dots$. Each
combination of sub-indices $a, i, j$ forms a single index $A = \{ aij \}$
of the matrix ${\cal M}_{AB}$. Since the possible values for the
sub-indices are $a = 1, 2$ and $i, j = 1, \dots N$, ${\cal M}$ is,
for now, a $2 N^2 \times 2 N^2$ matrix.

However, as it stands ${\cal M}$ has a pair of zero eigenvalues coming
from a zero mode of the fermionic action, which is invariant under a
change $\psi \rightarrow \psi + \epsilon \mathbbm{1}$. This zero mode
can be removed by omitting, in the partition function, the integration
over one of the diagonal elements of the $\psi$ matrices, for example
$\psi_{NN}$. This amounts to calculating the fermionic integral for a
sub-matrix of ${\cal M}$ in which the two rows and columns corresponding
to $i = j = N$ have been crossed out. Call the determinant of this
sub-matrix $\rm{det}_F(A)$. It is easily checked that $\rm{det}_F(A)$
is always real and positive semi-definite.

After integration over the fermions, the partition function becomes
\begin{equation}
Z = \int {\cal D}A \ e^{-S_B(A)} \ \rm{det}_F(A)
\end{equation}
where the integration measure is given by
\begin{equation}
{\cal D} A = \prod_\mu \
\left( \prod_{i>j} d {\rm Re} (A^\mu_{ij}) \ d {\rm Im} (A^\mu_{ij}) \right)
\left( \prod_{i} d A^\mu_{ii} \right) \ \delta({\rm Tr} A^\mu)
\label{measure}
\end{equation}
The most practical way to deal with the global constraints
${\rm Tr} A^\mu = 0$ in (\ref{measure}) is to just ignore them in the
simulations, taking $A^\mu$ to be arbitrary Hermitean matrices. This
creates a new zero mode in the bosonic sector, since the action is now
invariant under the change $A^\mu \rightarrow A^\mu + a^\mu \mathbbm{1}$,
which manifests as a random walk of ${\rm Tr} A^\mu$ in the simulations.
However, this can easily be corrected by simply subtracting the trace
from the matrices before each measuring step,
\begin{equation}
A^\mu_{ii} \rightarrow A^\mu_{ii} - \frac{1}{N} {\rm Tr} A^\mu \, .
\end{equation}
Physically, this means that we always take measurements in a reference
frame that is fixed to the system's center of mass. In the surface
theory, the shift corresponds to replacing
$X^\mu \rightarrow X^\mu - X^\mu_{CM}$.

To update the fields we use a standard Metropolis algorithm. Each updating
step consists of randomly choosing one element of each matrix $A^\mu$
and proposing a change by a random number taken from a uniform
distribution in a range $[-\epsilon, \epsilon]$, where $\epsilon$
is adjusted so as to produce reasonable acceptance rates. We accept or
reject the change according to the Metropolis criterion.

\section*{Results}

A first, easy test of our program consists of measuring the average
bosonic action, which we know from general scaling arguments should
behave as $\langle S_B \rangle = 3/2 (N^2 - 1)$.

\begin{figure}
\begin{center}
\psfrag{rho}{$\rho(\lambda)$}
\psfrag{lambda}{$\lambda$}
\includegraphics[height=8cm]{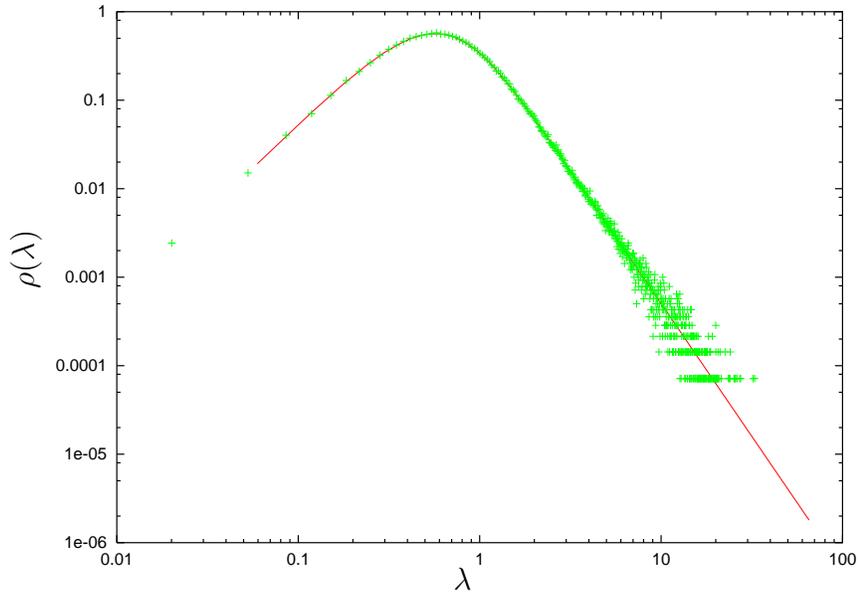}
\end{center}
\caption{\label{rhoN2}The distribution of eigenvalues from the
theoretical prediction (solid line) and the numerical data (crosses).
The scale is logarithmic on both axes.}
\end{figure}

Secondly, we compared the distribution of eigenvalues of $A^\mu$
to the theoretical formula for $N = 2$ \cite{su2},
\begin{equation}
\rho(\lambda) = 3 \sqrt{\frac{2}{\pi}} \lambda^2 \
U \left( \frac{5}{4}, \, \frac{1}{2}, \, 4\lambda^4 \right) \, ,
\end{equation}
where $U$ is the Kummer U-function.
The numerical results are shown, together
with the theoretical curve, in figure \ref{rhoN2}.

For large $\lambda$ (large $R^2$) the quality of the numerical data is
limited by two factors. For one thing, the number of data points in
this range is very small as the tail, although long, contains only a
small fraction of the whole distribution ($\approx 2.64\%$ for
$|\lambda| > 4$). Thus, it takes a long time for the algorithm to produce
reasonable statistics in this region. Secondly, the power of the tail
is, in absolute values, not very large, which means that once the
algorithm does enter the tail, it embarks on a random walk with long
excursions that increase the autocorrelation time enormously. Despite
these limitations, however, the figure shows quite good agreement
between the theoretical curve and the simulation data even within the tail.

Generally, it is known that for the reasons just given it is extremely
difficult to deal with power-like fall-offs in numerical simulations.
To improve the quality of the data in this region, we can set a lower
limit $R^2_{min}$ on $R^2$ to prevent the algorithm from going to the bulk of
the distribution, where it would spend almost all of its time otherwise.
The price to pay for such a cut-off is a decrease in the acceptance
rate of the algorithm as it frequently tries to push through this
boundary and go back to the main part. In practice, however, the drop
in the acceptance rate turns out to be not too severe. For example,
with $N = 4$ and $\epsilon = 0.1$ the acceptance rate decreases from
$74\%$ to $24\%$ after imposing a lower barrier of $R^2_{min} = 120$.

Similarly, to prevent the algorithm from making too long excursions
into the comparatively flat tail of the distribution we can introduce
an upper limit $R^2_{max}$. As expected, this drastically reduces the
autocorrelation time.

We measured the asymmetry parameter $\eta$ and the distribution of the
squared gyration radius $R^2$ for $N = 2, \ldots, 8$. For each value of $N$ 
we performed $10^5$ measurements
(or $4 \times 10^4$ in the case of $N = 8$), each separated by
100 sweeps, where a sweep encompasses $N^2$ Metropolis trials. As an
example, for $N = 4$, we found an integrated autocorrelation time for
$R^2$,  calculated in units of $100$ sweeps, of $\tau_{R^2} = 130(10)$.

Let us start with a discussion of the results from 
the $N = 2$ case. From the analytic 
solution (\cite{s,ss}), we find for the distribution
of the eigenvalues of $C_{\mu\nu}$~:
\begin{equation}
\rho (r) \sim (r_1 r_2 r_3)^\alpha
(r^2_1\!-\!r^2_2)(r^2_2\!-\!r^2_3)(r^2_3\!-\!r^2_1)
e^{-2 (r^2_1r^2_2+r^2_2r^2_3+r^2_3r^2_1)} \, .
\label{l4}
\end{equation}
All eigenvalues $r_i$, $i > 3$, are identically zero, {\em i.~e.} even
for small $R^2$ the system is only three-dimensional independently of $D$.
The exponent $\alpha$ depends on $D$ as $\alpha = 2D - 5$. The same formula,
with $\alpha = D - 3$, also describes the distribution
of eigenvalues of the purely bosonic model.

We can use equation (\ref{l4}) to explain the large $R$ behaviour of
the model, where in this case $R^2 = r^2_1 + r^2_2 + r^2_3$
(because $r^2_4 = 0$). When moving to large values of $R^2$, all
configurations will be exponentially suppressed, except for those that
keep the exponent
$r^2_1 r^2_2 + r^2_2 r^2_3 + r^2_3 r^2_1$
approximately constant.
Given that $R^2$ can only grow large if at least one of the eigenvalues
does so as well, this can be achieved only if
$r^2_1 \sim R^2$, $r^2_{2,3} \sim \frac{1}{R^2}$.
As a consequence we can, in \eqref{l4},
approximately replace $r_1 \rightarrow R$,
$r_{2,3} \rightarrow R^{-1}$ and
$\int d r_{2,3} \rightarrow R^{-1}$. This gives us
the distribution of $R$ in the large $R$ limit,
\begin{equation}
\rho (R) \sim R^{-\alpha} \, .
\end{equation}
\begin{figure}
\begin{center}
\psfrag{RR}{$R^2$}
\psfrag{r}{$\frac{r^2}{R^2}$}
\includegraphics[height=8cm]{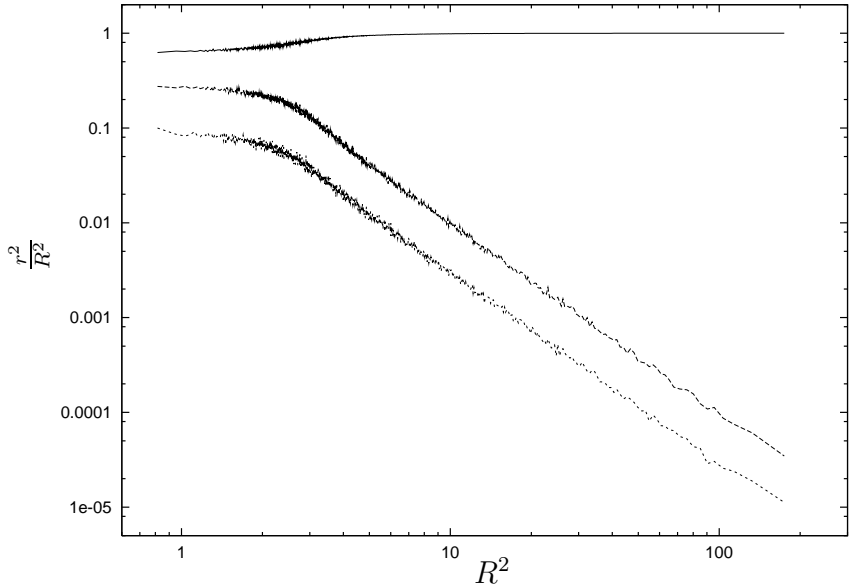}
\end{center}
\caption{\label{RR_r}The eigenvalues $r_1^2$, $r_2^2$, and $r_3^2$,
normalized with $R^2$, as
functions of $R^2$ for $N = 2$. Note that the data has been smoothed
so that each point in the figure actually represents a pair
$(\bar{R}^2, \bar{\eta})$, with the average taken over $n = 100$
successive values of $R^2$.}
\end{figure}
Since we have $\frac{r^2_{2,3}}{R^2} \sim \frac{1}{R^4}$ (see figure
\ref{RR_r}), we find for the asymmetry parameter
\begin{equation}
\eta \sim \frac{1}{(R^2)^2} \, ,
\end{equation}
{\em i.~e.} $\eta$ goes to zero for large $R^2$ and the quantum system
becomes one-dimensional.

To summarize for $N = 2$: The flat directions of the bosonic part of
the action, which correspond to constant values of the exponent in
(\ref{l4}), lead to a large $R$ behaviour of the eigenvalues
$r^2_1 \sim R^2$, $r^2_{2,3} \sim \frac{1}{R^2}$. This is true
independently of the presence or absence of fermions in the theory.
However, the addition of fermions does strongly influence the power
of the probability distribution $\rho(R)$.

Note that the flat directions being one-dimensional is a consequence
of the particular form of the bosonic action; we can create
systems of higher dimensions in the large $R$ range by choosing
a different action. For example, the action
$\rm{Tr} [A_\mu,A_\nu][A_\nu,A_\alpha][A_\alpha,A_\mu]$
has the same symmetries as $\rm{Tr} [A_\mu,A_\nu]^2$, but some of
its flat directions are now two-dimensional.
This can be seen from the eigenvalue distribution, which in this case reads
\begin{equation}
\rho (r) \sim (r_1 r_2 r_3)^\alpha
(r^2_1\!-\!r^2_2)(r^2_2\!-\!r^2_3)(r^2_3\!-\!r^2_1)
e^{-24 r_1^2r^2_2r^2_3} \, .
\end{equation}
For large $R$, one can keep the exponent 
$r_1^2 r^2_2 r^3_2$ constant 
by choosing $r_1 \sim R$, $r_2 \sim R$, 
and $r_3 \sim 1/R$, which obviously describes a two-dimensional disc.
In this case $\eta(R)$ goes toward a constant for large $R$.
This type of asymmetry could be traced by a higher dimensional
counter-term of $\eta$, $\eta_2 = 1 - \frac{r_1^2+r_2^2}{R^2}$,
which would vanish for large $R$.

\begin{figure}
\begin{center}
\psfrag{eta}{$\eta$}
\psfrag{RR}{$R^2$}
\includegraphics[height=8cm]{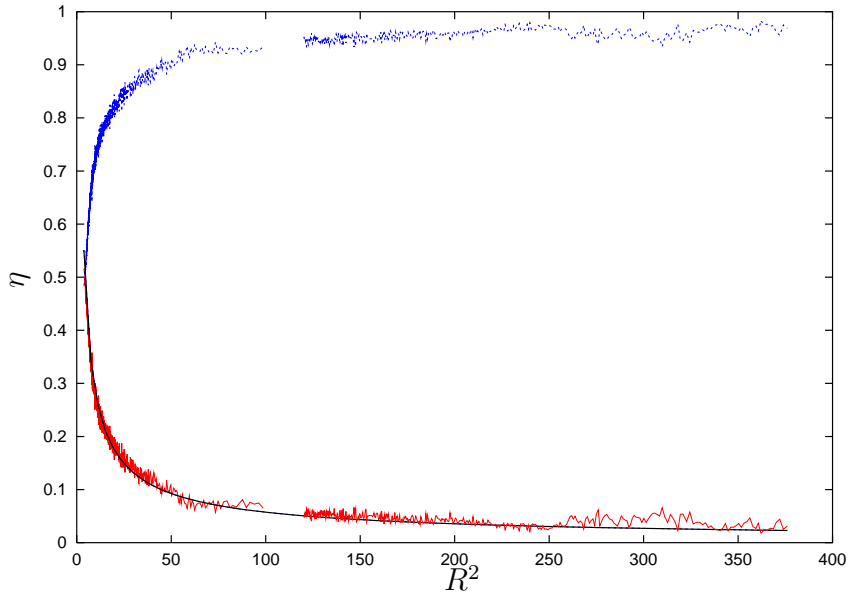}
\end{center}
\caption{The plot shows $\frac{r^2_1}{R^2}$ (upper curve)
and $\eta = \frac{r^2_2 + r^2_3 + r^2_3}{R^2}$ (lower curve)
as functions of $R^2$, for $N = 4$. The data has been smoothed as
explained for figure \ref{RR_r}. 
The plot combines two data sets, one from an unconstrained simulation
(the points for $R^2 < 100$) and one from a simulation with a lower
cut-off $R^2_{min} = 120$. The solid line is the best power-like fit
to the data from the unconstrained simulation.}
\label{eta}
\end{figure}

Coming back to the standard model, we can now use Monte Carlo simulations
to repeat the same analysis for
$N \ge 3$. Figure \ref{eta} shows the numerical data for $\eta$ measured
as a function of $R^2$ for $N = 4$. As for $N = 2$, for large $R^2$,
$\eta$ vanishes -- in other words, the main contribution to the square
extent comes from the elongated direction of what is thus an
essentially one-dimensional system.

We fitted the data to a power-law function $\eta = a (R^2)^b$, and found
for $N = 4$ as the best fit $a = 1.378(7)$ and $b = -0.690(4)$.
To test the stability of this result in the region of large $R^2$, we
repeated the simulation with the inclusion of a lower bound
$R^2_{min} = 120$ as discussed in the previous section, 
to see whether the validity of the fit
goes beyond the range of $R^2$ generated in the first run. The results
are collected in figure \ref{eta}, and do indeed show good agreement
with the results of the first simulation.

Finally, we collect the results of the best power-law fits to $\eta$
for $N = 3, 4, 5, 6, 8$ in the following table. For comparison, we also
include the results for the surface model with $n_T = 8, 12, 20, 28$
triangles.

\begin{center}
\begin{tabular}{|c|c||c|c|}
\hline
$N$ & $b_{matrix}$ & $n_T$ & $b_{surface}$ \\
\hline
3   & -0.825(2)    & 8     & -1.604(16)    \\
4   & -0.690(4)    & 12    & -1.463(19)    \\
5   & -0.606(4)    & 20    & -1.412(15)    \\
6   & -0.562(5)    & 28    & -1.383(20)    \\
8   & -0.503(29)   &       &               \\
\hline
\end{tabular}
\end{center}
In both the matrix and the surface model, the exponent 
$b$ changes with $N$. If we assume the finite size corrections to be
of order $\frac{1}{N}$, that is $b (N) = b_\infty + \frac{c}{N}$,
we can try to estimate $b_\infty$ from a fit to the measured values.
For the matrix model, this leads to $b_\infty = -0.35(1)$, $c = -1.37(4)$.
We also see that $b_\infty$ stays negative and far from zero if we assume
a different kind of finite size correction, like $\frac{1}{N^2}$ or
$\frac{1}{\sqrt{N}}$. Thus, the data suggests that the dominance of
one-dimensional configurations in the large $R$ part of the spectrum
should also be present in the limit of $N \rightarrow \infty$. To answer
this question conclusively, however, one should extend the simulations
to larger values of $N$.

Another related issue is 
the scaling of the gyration radius with $N$. From results 
of the one-loop approximation \cite{1loop}, one expects
the maximum of the gyration radius distribution to scale as
$\sqrt{R^2_{max}} (N) \sim N^{1/4}$, corresponding to the
Hausdorff dimension $d_H = 4$ of branched polymers which appear naturally
in this approximation. Indeed, if we fit our numerical results for the
surface model to the formula $\sqrt{R^2_{max}} (N) = a N^{b}$,
we find $a = 0.573(4)$ and $b = 0.234(4)$, which is close to $1/4$ even
though the surfaces we studied are quite small ($n_T = 12, \dots, 60$).
Alternatively, we can do the same using another quantity $\widehat{R}$,
which was defined in \cite{mc2} as an alternative estimate of the system
extent~:
\begin{equation}
\widehat{R} = \frac{1}{N} \sum_i |X_i - X_{CM}| \, .
\end{equation}
Here, the best power-like fit gives $a = 0.579(6)$ and $b = 0.204(4)$,
which does not differ too much from the values obtained for the generic
definition of the gyration radius. Presumably, for larger values of $N$
we should see $b$ converging to $1/4$ for either definition.

In the matrix model, on the other hand, the situation looks quite different
in the simulated range of $N$. Namely, in this case the best fit to
$\sqrt{R^2_{max}} (N) = a N^b$ is $a = 1.03(4)$ and $b = 0.81(6)$,
which means $d_H = \frac{1}{b} = 1.23(9)$. This is far away from the
result of the one-loop approximation, $d_H = 4$. However, one should note
that the simulated systems are rather small for branched polymers to fully
develop. Indeed, in \cite{mc2} a semi-classical analogon of $\widehat{R}$
constructed from commuting coordinates was studied for values of $N$ in
the range $N = 16, \dots, 48$, and its average was estimated to behave as
$\langle \widehat{R}_{sc} \rangle_N \sim N^{1/4}$, which should 
also mean $\langle \sqrt{R^2} \rangle_N \sim N^{1/4}$.

Let us close this section with a remark about the definition of
dimensionality of the system. There are two natural possibilities~: the
Hausdorff dimension $d_H$, and the effective dimension coming from the
principal component analysis of the correlation matrix $C_{\mu\nu}$.
The former describes the relation between the system size and its average
extent, whereas the latter gives the minimal dimensionality of the subspace 
to which we can restrict our description of the system without neglecting
important degrees of freedom. The two may differ in general, and it is
an important physical question which definition should be used for
any given problem \cite{ssb}.

\section*{Conclusions}

We have shown that the underlying geometry that can be associated
with the field configurations of reduced supersymmetric Yang-Mills
integrals in $D = 4$ dimensions shrinks in the limit of large
$R^2 = \frac{1}{N} \, {\rm Tr} \, A_\mu^2$ to an essentially
one-dimensional tube. The original rotational symmetry
of the action is spontaneously broken to the direction of the tube. 
The same mechanism has already been observed 
in numerical simulations of the corresponding surface model.

The origin of the power-like behaviour 
boils down to the existence of flat directions in the bosonic 
part of the action. The tubes correspond to field configurations 
that expand along the valleys of these flat 
directions, where they do not have to pay the usual
`exponential price', but rather one that is only power-like. The exact
value of this power comes from the prefactor, which depends on the
dimensionality of the problem. However, it seems that the power does not
change with $N$. For $D = 4$, the distribution of the gyration radius
was determined as $\rho(R) \sim R^{-3}$.

The analysis presented in this paper should be applicable to the
six- and ten-dimensional cases as well, where one expects a
distribution $\rho(R) \sim R^{-2D+5}$. In the large $N$ limit we
expect the estimator of the linear system extent as defined by
$R_k \equiv \langle R^{2k} \rangle^{\frac{1}{2k}}$ to pick up
the singularity for $k$ large enough, which may lead to a different
scaling with $N$ and thus a different Hausdorff dimension. A similar
thing happens for example with L\'evy random walks, which are known
to have a Hausdorff dimension smaller than two \cite{bg}. For the
IKKT model in particular, an eigenvalue distribution with a power-like
tail as the one suggested here would mean that we should expect
higher-order correlators starting with $k \ge 7$ to effectively
sample the tail of the distribution, which is dominated by
one-dimensional configurations.

\section*{Acknowledgments}

We thank Jan Ambj\o{}rn, Piotr Bialas, Jun Nishimura and Matthias Staudacher
for discussions. The work was partially supported by the EC IHP network
{\it HPRN-CT-1999-000161} and the KBN grant 2P03B00814.


\begin{thebibliography}{99}

\bibitem{ek}
  T. Eguchi and H. Kawai, \
    Phys. Rev. Lett. {\bf 48} (1982) 1063.

\bibitem{gk}
  D. J. Gross and Y. Kitazawa, \
    Nucl. Phys. {\bf B206} (1982) 440.

\bibitem{s}
  A. V. Smilga, \
    Nucl. Phys. {\bf B266} (1986) 45.

\bibitem{y}
  P. Yi, \
    Nucl. Phys. {\bf B505} (1997) 307.

\bibitem{ss}
  S. Sethi and M. Stern, \
    Comm. Math. Phys. {\bf 194} (1998) 675.

\bibitem{ikkt}
  N. Ishibashi, H. Kawai, Y. Kitazawa and A. Tsuchiya, \
    Nucl. Phys. {\bf B498} (1997) 467.

\bibitem{su2}
  W. Krauth, J. Plefka and  M. Staudacher, \ 
    Class. Quant. Grav. {\bf 17} (2000) 1171-1179.

\bibitem{1loop}
  H. Aoki, S. Iso, H. Kawai, Y. Kitazawa and A. Tada, \
    Prog. Theor. Phys. {\bf 99} (1998) 713.

\bibitem{mc1}
  W. Krauth, H. Nicolai and M. Staudacher, \
    Phys. Lett. {\bf B431} (1998) 31.

\bibitem{tail}
  W. Krauth and M. Staudacher, \
    Phys. Lett. {\bf B453} (1999) 253.

\bibitem{mc2}
  J. Ambj\o{}rn, K. N. Anagnostopoulos, W. Bietenholz, T. Hotta and
  J. Nishimura, \
    JHEP {\bf 0007} (2000) 013.

\bibitem{mc2'}
  J. Ambj\o{}rn, K. N. Anagnostopoulos, W. Bietenholz, T. Hotta and
  J. Nishimura, \
    {\tt hep-lat/0009030}.

\bibitem{phase}
  J. Ambj\o{}rn, K. N. Anagnostopoulos, W. Bietenholz, T. Hotta and
  J. Nishimura, \
    JHEP {\bf 0007} (2000) 011.

\bibitem{largeN}
  I. Bars, \
    Phys. Lett. {\bf B245} (1990) 35.

\bibitem{largeN'}
  M. Fukuma, H. Kawai, Y. Kitazawa and A. Tsuchiya, \ 
    Nucl. Phys. {\bf B510} (1998) 158.

\bibitem{oy}
  S. Oda and T. Yukawa, \
    Prog. Theor. Phys. {\bf 102} (1999) 215.

\bibitem{bbpt}
  P. Bialas, Z. Burda, B. Petersson and J. Tabaczek, \
    Nucl. Phys. {\bf B591} (2000) 1.

\bibitem{ssb}
  J. Nishimura and G. Vernizzi, \
    JHEP {\bf 0004} (2000) 015.

\bibitem{bg} J.-P. Bouchaud and A. Georges, \
    Phys. Rep. {\bf 195} (1990) 127.

\end{thebibliography}
\end{document}